\documentclass[runningheads]{llncs}
\usepackage{graphicx}
\usepackage[utf8]{inputenc}
\usepackage{todonotes}
\usepackage{xcolor}
\usepackage{soul}
\usepackage{url}
\usepackage{hyperref}

\definecolor{greenish}{HTML}{a1d99b}

\begin{document}
\title{Overview of LiLAS 2020 -- Living Labs for Academic Search}
%
%

\author{Philipp Schaer\inst{1}\orcidID{0000-0002-8817-4632} 
\and
Johann Schaible\inst{2}\orcidID{0000-0002-5441-7640} 
\and
Leyla Jael Garcia Castro\inst{3}\orcidID{0000-0003-3986-0510}
}
\authorrunning{Schaer, Schaible, Garcia Castro}
%
\institute{
TH Köln - University of Applied Sciences, Germany\\ 
\email{philipp.schaer@th-koeln.de}\\
\and
GESIS - Leibniz Institute for the Social Sciences, Germany
\email{johann.schaible@gesis.org}\\ 
\and
ZB MED - Information Centre for Life Sciences, Germany\\
\email{ljgarcia@zbmed.de}
}
\maketitle              
\begin{abstract}

Academic Search is a timeless challenge that the field of Information Retrieval has been dealing with for many years. Even today, the search for academic material is a broad field of research that recently started working on problems like the COVID-19 pandemic. However, test collections and specialized data sets like CORD-19 only allow for system-oriented experiments, while the evaluation of algorithms in real-world environments is only available to researchers from industry. In LiLAS, we open up two academic search platforms to allow participating research to evaluate their systems in a Docker-based research environment. This overview paper describes the motivation, infrastructure, and two systems LIVIVO and GESIS Search that are part of this CLEF lab.


\keywords{Evaluation, living labs, academic search, reproducibility}
\end{abstract}
%
%
%
\section{Introduction}

The field of Information Retrieval (IR) originated in the domain of scientific/academic information and documentation. Back in the 1960s, the original Cranfield studies dealt with the indexation and the retrieval of scientific documents. Cleverdon et al. established their whole evaluation methodology around the use-case of scientific and academic retrieval requirements. Today, the search for relevant scientific documents is still an open endeavor, and although retrieval systems show substantial performance gains, it is not a solved problem yet. The current COVID-19 pandemic, for example, showed once again that even old problems like the search for scientific documents are not solved. Therefore, current efforts like the CORD-19 collection\footnote{\url{https://www.semanticscholar.org/cord19}} and the TREC COVID retrieval campaign\footnote{\url{https://ir.nist.gov/covidSubmit/index.html}} gather much attraction and are in the spotlight of the IR community, even though at their core, they deal with the same -- timeless -- problem set as Cleverdon more than 50 years ago. 


Besides these timeless retrieval issues, the need for innovation in academic search is shown by the stagnating system performance in controlled evaluation campaigns, as demonstrated in TREC and CLEF meta-evaluation studies \cite{yang_critically_2019,armstrong_improvements_2009}. User studies in real systems of scientific information and digital libraries show similar conditions. Although massive data collections of scientific documents are available in platforms like arXiv, PubMed, or other digital libraries, central user needs and requirements remain unsatisfied. The central mission is to find both relevant and high-quality documents - if possible, directly on the first result page. Besides this ad-hoc retrieval problem, other tasks such as the recommendation of relevant cross-modality content including research data sets or specialized tasks like expert finding are not even considered here. On top of that, relevance in academic search is multi-layered \cite{DBLP:conf/ecir/CarevicS14} and a topic that drives research communities like the Bibliometrics-enhanced Information Retrieval (BIR) workshops \cite{DBLP:conf/ecir/MayrSLSM14}.


The Living Labs for Academic Search (LiLAS) workshop fosters the discussion, research, and evaluation of academic search systems, and it employs the concept of living labs to the domain of academic search~\cite{Schaible2020}. The goal is to expand the knowledge on improving the search for academic resources like literature, research data, and the interlinking between these resources. To support this goal, LiLAS introduces an online evaluation infrastructure that directly connects to real-world academic search systems~\cite{schaer_ecir2019}. LiLAS cooperates with two academic search systems providers from Life Sciences and Social Sciences. Both system providers support LiLAS by allowing participants of the lab to employ experimental search components into their production online system. We will have access to the click logs of these systems and use them to employ A/B tests or more complex interleaving experiments. Our living lab platform STELLA makes this possible by bringing platform operators and researchers together and providing a methodological and technical framework for online experiments~\cite{breuer2019stella}. 

\section{Related Work from CLEF and TREC}

CLEF and TREC hosted the Living Labs for Information Retrieval (LL4IR) and Open Search (TREC-OS) initiatives that are the predecessors to LiLAS. Both initiatives shared a common evaluation infrastructure that was released as an API\footnote{\url{https://bitbucket.org/living-labs/ll-api}}. This API allows academic researchers to access the search systems of other platforms. Participants of LL4IR and TREC-OS had access to the search systems' head queries and document sets. They had to precompute ranked result lists for a given set of candidate documents for the given head queries. Therefore it was a typical ad-hoc search task. Another task was run during the CLEF NewsReel campaign, where participants had to recommend news articles. This was possible by employing an offline test collection or in real-time via the Open Recommendation Platform (ORP) used by PLISTA. 

All these labs can be considered living labs and represent a user-centric study methodology for researchers to evaluate retrieval systems' performance within real-world applications. Thus, they aim to offer a more realistic experiment and evaluation environment as offline test collections and therefore, should be further investigated to raise IR-evaluation to a more holistic level. 

Within TREC and CLEF, only very few tracks or labs focused on the evaluation of academic search systems. Some used scientific documents or use-cases to generate test collections but did not necessarily focus on the unique requirements of the academic domain. Within CLEF, the Domain-specific track~\cite{kluck_domain-specific_2000} compiled a collection of bibliographic records and research project descriptions from the Social Sciences to test the needs of scientific retrieval tasks. This test collection was created to contrast the then usual ``general-purpose news documents'' and to employ ``different search criteria than those used for reference retrieval in databases of scientific literature items, and also offer no possibility for comparable test runs with domain-specific terminology''.
More recently, the TREC Precision Medicine / Clinical Decision Support Track released a large test collection in 2016 based on open access full-text documents from PubMedCentral. TREC-COVID is the latest retrieval campaign aiming at academic search with a particular focus on the rapidly growing corpus of scientific work on the current COVID-19 pandemic.

The LiLAS workshop is a blend of the most successful parts of these previous evaluation campaigns. 
The Domain-specific track had a strong focus on scientific search, thesauri, and multilingual search. NewsREEL had an active technological component, and LL4IR/TREC-OS turned from product search to academic search but was not able to implement the scientific focus into the last iteration. There is much potential that is still not used in the question of how to evaluate academic search platforms online.

\section{STELLA -- Evaluation Infrastructure for LiLAS}
\label{sec:infrastructure}

Nowadays, testing approaches are commonly used to try out and evaluate how users interact when presented with some new or modified features on a website. Whenever the new or modified features differ from what has been done before, when multiple features change at once, or when the user interaction is to be gathered in a systematic way, A/B testing comes into place. An A/B testing, a controlled online experiment, allows to expose a percentage of real users and life-test those new or modified features \cite{bakshy_designing_2014,knijnenburg_conducting_2012} offering to website designers and developers a living lab where to assess reactions and usage of new features better, allowing them for more accurate tuning based on data collected from production systems.

\begin{figure}[t]
    \centering
    \includegraphics[width=\linewidth]{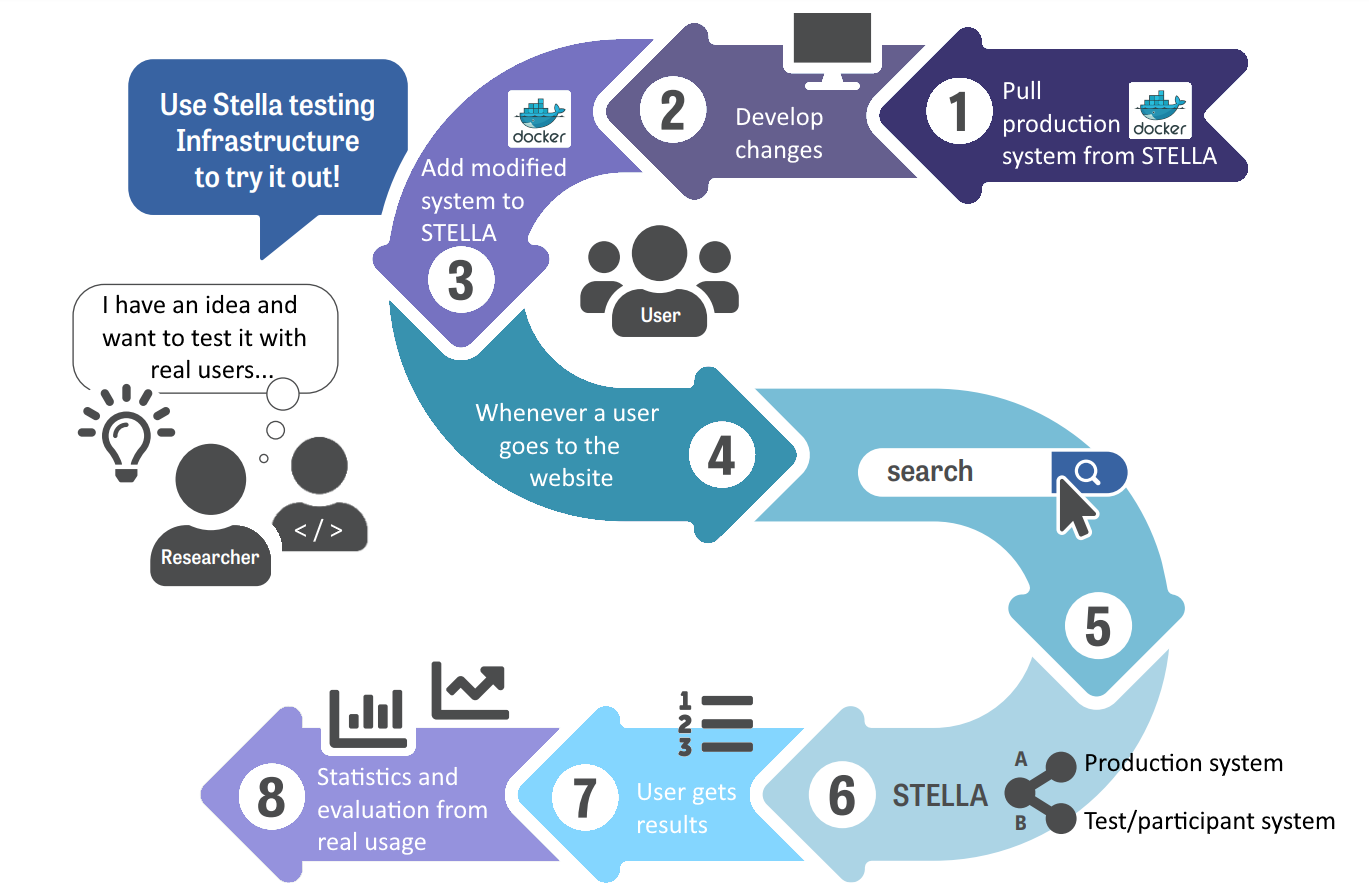}
    \caption{\textbf{STELLA workflow, an online living lab supporting testing from ideas to evaluation}: Participants package their systems with the help of Docker containers that are deployed in the backend of academic information retrieval and recommendation systems. Users interact directly with the system, with a percentage diverted to the experimental features. Researchers and developers retrieve results and deliver feedback to tune and improve changes.}
    \label{fig:infrastructure}
\end{figure}

For LiLAS, we use STELLA as our living lab evaluation infrastructure. STELLA is aiming to make it easier to evaluate academic retrieval information and recommendation systems \cite{breuer2019stella}. Figure \ref{fig:infrastructure} shows an overview of how the steps flow from a researcher's or developer's idea to the evaluation feedback so the changes can be tuned and improved. It all starts with an idea, for instance adding synonyms to the keywords used by an end-user when searching for information. Developers will work on a modified version of the production system, including this change they want to analyze. Whenever an end-user goes to the system, everything will look as usual. Once the search keywords are introduced, STELLA will show end-user some results from the experimental system and some results from the regular production system. End-users will continue their regular interaction with the system. Based on the retrieved documents and the following interaction, STELLA will create an evaluation profile together with some statistics. Researchers and developers will then analyze STELLA's feedback and will react accordingly to get the usage level they are aiming at.

STELLA's infrastructure relies on the container virtualization environment Docker \cite{merkel_docker_2014}, making it easier for STELLA to run multiple experimental systems, i.e., a multi-container environment, and compare them to each other and the production system as well. The core component in STELLA is a central Application Public Interface (API) connecting data and content providers with experimental systems, aka participant systems or participants, encapsulated as Docker containers. Further information can be found at the project website\footnote{\url{https://stella-project.org/}}, including some technical details via a series of blogs published regularly.

Currently, STELLA supports two main tasks: ad-hoc retrieval and recommendation. In the following subsections, we will introduce two systems used during the STELLA development phase to understand better, learn, and test these two tasks. Although a fully functional version is already available, there is still room for improvement, particularly regarding the events logging, statistics analysis, and overall evaluation. LiLAS will promote an early discussion with future adopters and participants that will benefit not only STELLA but living labs in general.

\subsection{LIVIVO}

\begin{figure}[t]
    \centering
    \includegraphics[width=1\linewidth]{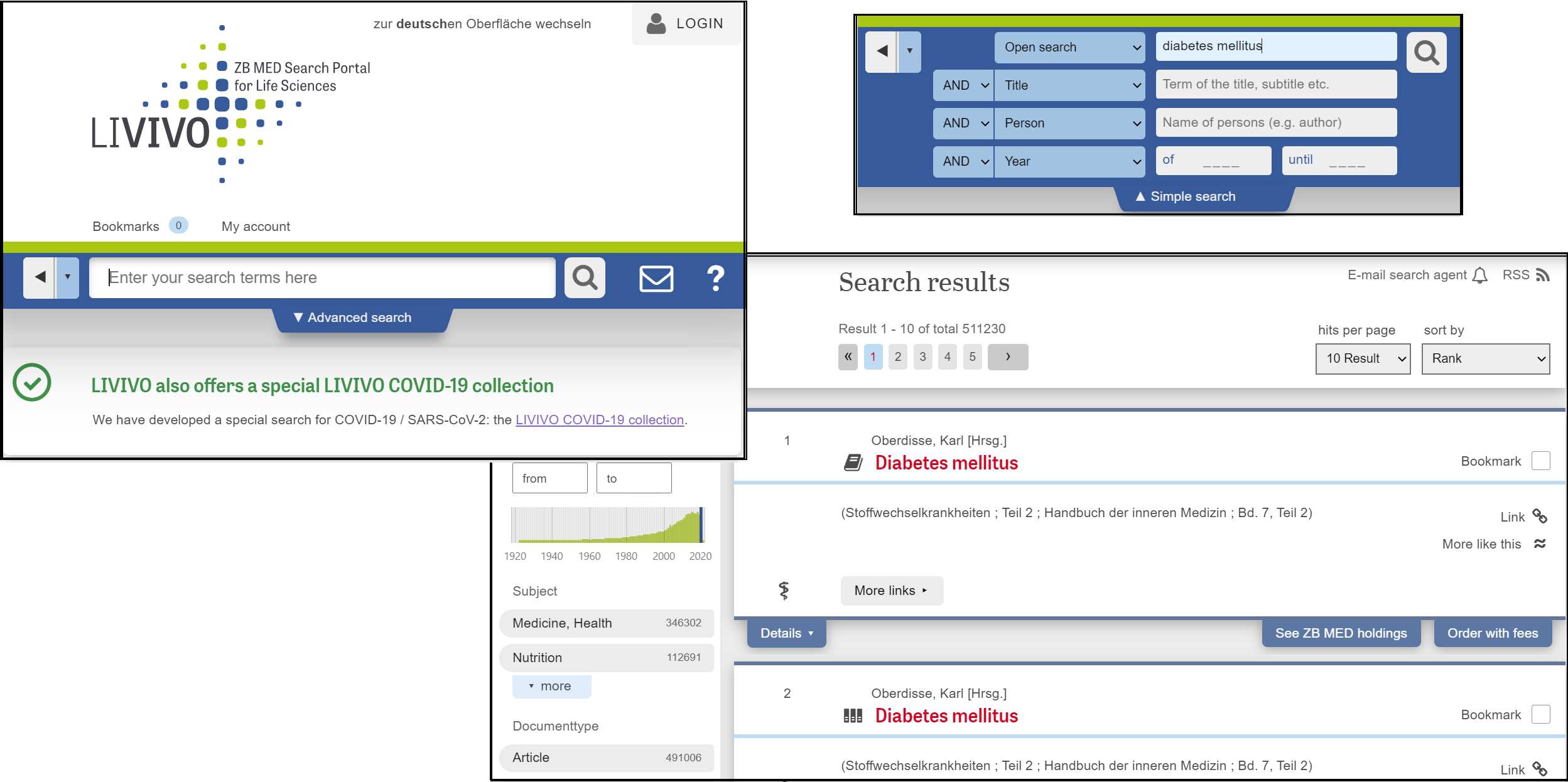}
    \caption{LIVIVO, the ZB MED retrieval platform. Users can search by keywords, title, author or year and obtain a result set sorted by relevance together with additional features such is filters, publication details, access links, other publications like the one on display and library stock.}
    \label{fig:livivo}
\end{figure}

LIVIVO\footnote{\url{https://livivo.de}} is a retrieval platform provided by ZB MED -- Information Centre for Life Sciences. It serves the Life Sciences domain with a focus on medicine, health, nutrition,  environment, and agriculture. LIVIVO includes unique features tailored to the German public and the national inter-library loan system, making it easier for researchers, practitioners, students, and the general public to access material licensed and hosted at different German libraries. LIVIVO brings together publication from 30 different sources, e.g., Medline, AGRICOLA, and AGRIS, including more than 58 million publications in different languages including English, German, Spanish, French and Portuguese. It uses automatic and semantic links to well-known vocabularies, for instance the Medical Sub-heading (MeSH) \cite{national_library_of_medicine_medical_2020} for medical sciences, UMTHES \cite{fock_environmental_2009} for environmental sciences, and AGROVOC \cite{caracciolo_agrovoc_2013} for agricultural sciences. The resultset ranked by relevance and can be narrowed down using filters such as the ZB MED subject fields. We include a sample query and corresponding results in \ref{fig:livivo}. From March 2020, there is a dedicated portal serving Covid-19 related information.

Regarding the integration to STELLA, a test instance of LIVIVO has been set up with a twofold purpose: introducing those elements needed in LIVIVO to integrate it to the STELLA framework, e.g., calling the STELLA API whenever a search is triggered in the production system, and evaluating the STELLA framework itself, i.e., how the containerization. Communication via the API and central STELLA server work with real production systems. We are also working on a LIVIVO dataset suitable for participant systems, mainly targeting MEDLINE articles written in English, about 25 million abstracts with their corresponding metadata including title, authors, affiliations, MeSH term among others.

\subsection{GESIS Search}

\begin{figure}[t]
    \centering
    \includegraphics[width=0.49\linewidth]{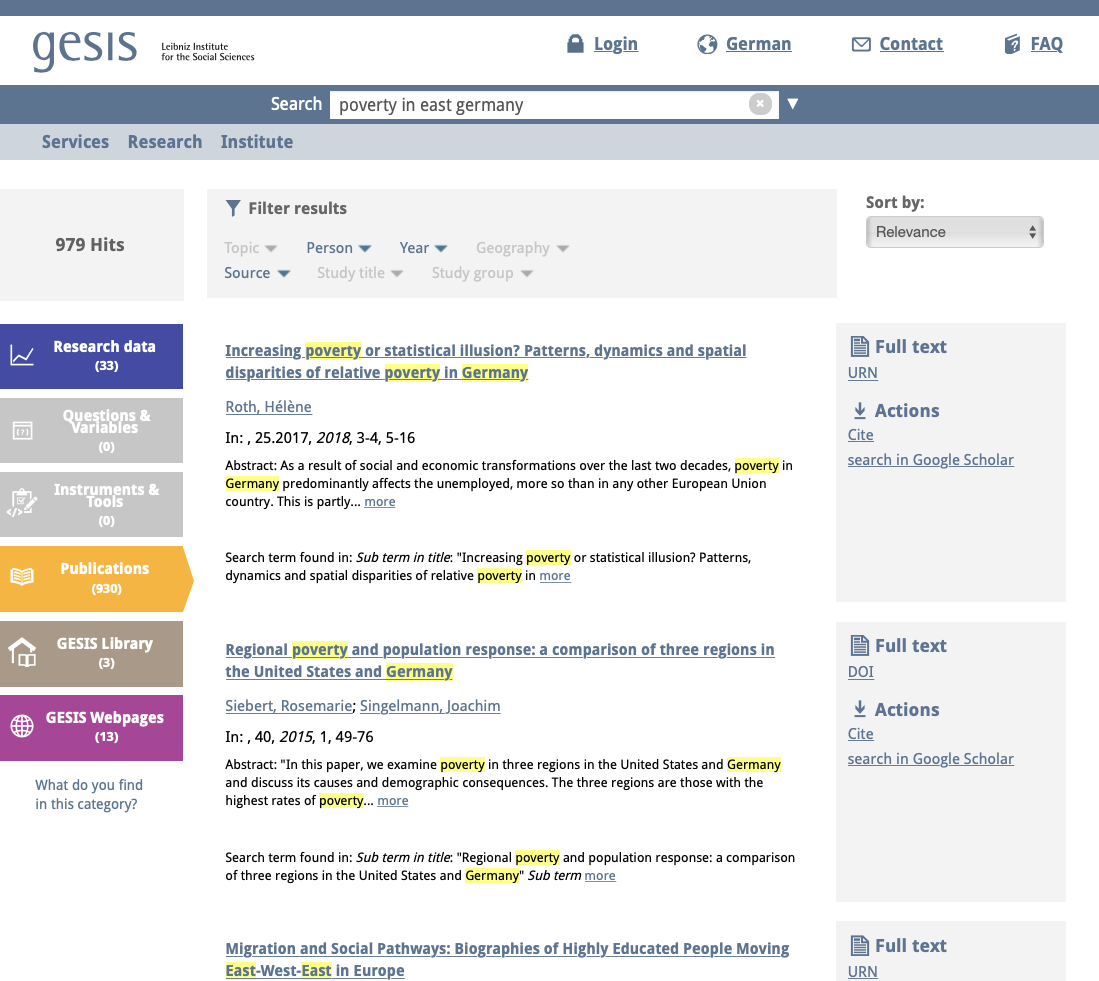}
    \includegraphics[width=0.49\linewidth]{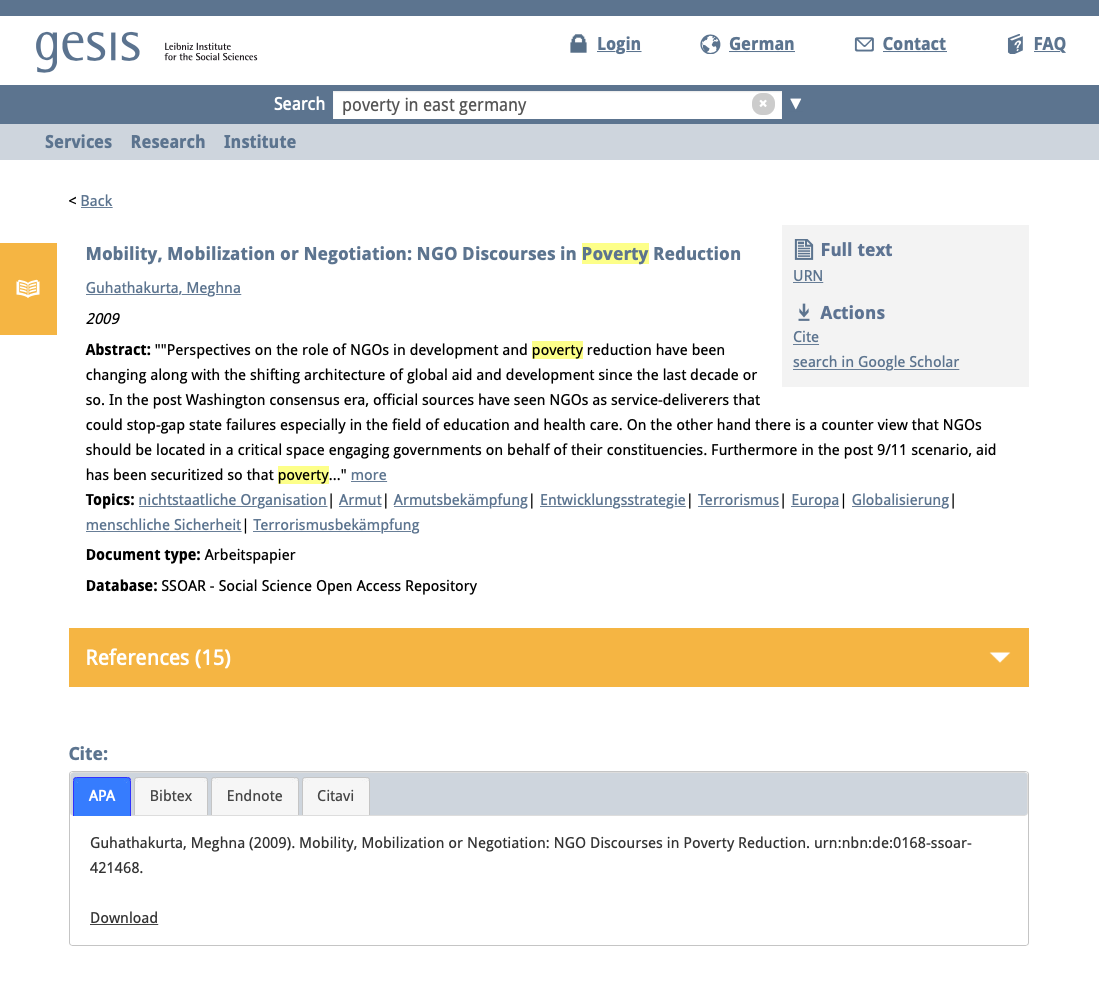}
    \caption{GESIS Search, social science information portal. It provides different types of information from the social sciences, like literature, research data, questions and variables, as well as instruments and tools. }
    \label{fig:my_label}
\end{figure}



The internal GESIS academic search GESIS Search\footnote{\url{https://search.gesis.org/}} aims to aid their users in finding appropriate scholarly information on the broad topic of social sciences~\cite{DBLP:conf/jcdl/HienertKBZM19}. To this end, it provides different types of information from the social sciences, comprising literature (95k publications), research data (84k), questions and variables (12.7k), as well as instruments and tools (370). 
The publications are mostly in English and German and are annotated with further textual metadata like title, abstract, topic, persons, and others. With the Social Science Open Access Repository (SSOAR)\footnote{\url{https://www.gesis.org/en/ssoar/home}}, GESIS Search also provides access to nearly 60k open access publications. Metadata on research data comprises (among others) a title, topics, datatype, abstract, collection method, primary investigators, and contributors in English and/or German. 

Regarding STELLA, the amount of different types of data allows not only for typical recommendations, such as from publications to publications but also for \emph{cross-domain recommendations}, i.e., recommendations from different types such as from publications to research data. While this is still work in progress, the GESIS Search data and possible relevance indicators, such as click-paths, can be obtained. The data can be used to train a recommender and report lessons learned, and file issue requests on how to improve the training data.  

\section{Conclusion and Outlook}

We presented the artifacts we would like to use for the actual evaluation tasks at CLEF 2021. These artifacts are: (1) the STELLA living lab evaluation infrastructure, and (2) the two academic search systems LIVIVO and GESIS Search. These systems are from the two disjunct scientific domains life sciences and social sciences and include different metadata on research articles, data sets, and many other entities. 

With this at hand, we will derive the CLEF 2020 workshop participants' evaluation tasks for CLEF 2021. Promising task candidates are:

\begin{itemize}
    \item Ad-hoc retrieval for life Science documents
    \item Dataset recommendation
\end{itemize}
\noindent
These tasks allows us to use the different data types available in the platforms. 

\section*{Acknowledgements}
This work was partially funded by the German Research Foundation (DFG) under the project no. 407518790.

\bibliographystyle{splncs04}
\bibliography{clef-lncs-2020}

\end{document}